\documentclass[aps,twocolumn,amsmath,amssymb,reprint]{revtex4}

\usepackage{graphicx}
\usepackage{dcolumn}
\usepackage{bm}

\usepackage[utf8]{inputenc}
\usepackage[T1]{fontenc}
\usepackage{mathptmx}

\usepackage{url,lineno,microtype}

\usepackage{tabularx}
\usepackage{dcolumn}
\usepackage{multirow}
\usepackage{booktabs}
\usepackage{latexsym}       
\usepackage[super]{nth}     

\begin{document}

\preprint{AIP/123-QED}

\title[Machine Learning assisted Chimera and Solitary states in Networks]{Machine Learning assisted Chimera and Solitary states in Networks}

\author{Niraj Kushwaha}
\author{Naveen Kumar Mendola}%
\affiliation{ 
Complex Systems Lab, Discipline of Physics, Indian Institute of Technology Indore, Khandwa Road, Simrol, Indore-453552, India
}%

\author{Saptarshi Ghosh}
\affiliation{ 
Complex Systems Lab, Discipline of Physics, Indian Institute of Technology Indore, Khandwa Road, Simrol, Indore-453552, India
}%

\author{Ajay Deep Kachhvah}
\affiliation{ 
Complex Systems Lab, Discipline of Physics, Indian Institute of Technology Indore, Khandwa Road, Simrol, Indore-453552, India
}%

\author{Sarika Jalan}
\affiliation{ 
Complex Systems Lab, Discipline of Physics, Indian Institute of Technology Indore, Khandwa Road, Simrol, Indore-453552, India
}%
\affiliation{ 
Discipline of Biosciences and Biomedical Engineering, Indian Institute of Technology Indore, Khandwa Road, Simrol, Indore-453552, India
}%
\affiliation{ 
Center for Theoretical Physics of Complex Systems, Institute for Basic Science(IBS), Daejeon 34126, Korea
}%
\email{sarikajalan9@gmail.com}

\date{\today}

\begin{abstract}
Chimera and Solitary states have captivated scientists and engineers due to their peculiar dynamical states corresponding to the co-existence of coherent and incoherent dynamical evolution in coupled units in various natural and artificial systems. It has been further demonstrated that such states can be engineered in systems of coupled oscillators by the suitable implementation of communication delays. Here, using supervised machine learning, we predict (a) the precise value of delay which is sufficient for engineering chimera and solitary states for a given set of system parameters, as well as (b) the intensity of incoherence for such engineered states. The results are demonstrated for two different examples consisting of single layer and multi layer networks. First, the chimera states (solitary states) are engineered by establishing delays in the neighboring links of a node (the interlayer links) in a 2-D lattice (multiplex network) of oscillators. Then, different machine learning classifiers, KNN, SVM and MLP-Neural Network are employed by feeding the data obtained from the network models. Once a machine learning model is trained using a limited amount of data, it makes predictions for a given unknown systems parameter values. Testing accuracy, sensitivity, and specificity analysis reveal that MLP-NN classifier is better suited than Knn or SVM classifier for the predictions of parameters values for engineered chimera and solitary states. The technique provides an easy methodology to predict critical delay values as well as the intensity of incoherence for designing an experimental setup to create solitary and chimera states. 
\end{abstract}

\maketitle

\section{Introduction}
\noindent In the year 2002, a new area was introduced in the field of nonlinear dynamics when Kuramoto {\it et al.} brought to light the phenomenon of occurrence of symmetry breaking in a system of identically coupled oscillators \cite{1_Chimera_main_paper}. Apart from synchronous and asynchronous states, they identified a remarkable hybrid dynamical structure where both the asynchronous and synchronous regions coexisted in a system of identical oscillators. Later, this mixed state of coherence and incoherence was termed as ‘chimera’, coined by Strogatz and Abrams \cite{2_Abrams_Strogatz_Chimera}. Initially identified in a system of identical Kuramoto oscillators, the chimera state has been pinpointed in a variety of other network models such as FitzHugh-Nagumo oscillators \cite{Semenova2016, Soli_FN1}, R\"{o}ssler oscillators~\cite{Shepelev2018}, van der Pol oscillators~\cite{Bukh2019}, coupled Rulkov maps~\cite{Rybalova2019b}, coupled maps \cite{Chaotic_map1}, coupled chaotic oscillators~\cite{Anishchenko2019}, multi-layer neuronal models \cite{majhi2016chimera}, Morris-Lecar neurons \cite{Calim2018},  modular neural network \cite{Hizanidis2016},  neuronal network model of the cat brain~\cite{Santos2017} and data-driven model of the brain \cite{Bansal2019}. Over the years, the researchers have spotted similar fascinating chimeric patterns and labeled them as virtual chimera \cite{3_Virtual_Chimera}, traveling chimera \cite{4_Traveling_Chimera}, breathing chimera \cite{5_Breathing_Chimera}, spike-burst chimera states~\cite{Santos2019} and others \cite{7_, 8_}. Several approaches were made to provide an analytical explanation for the emergent chimera state~\cite{omel2018mathematics, rakshit2019transitions}. A comprehensive review on the development of chimera states in a variety of systems can be found in ~\cite{bera2017chimera, majhi2019chimera}.

Recently, another chimera-like pattern, the solitary states, has attracted tremendous attention of the scientific community. The word solitary originated from Latin `solitarius' stands for `alone' or `isolated'. In solitary states, unlike chimeric patterns, a few identical oscillators are split off from different isolated locations in the synchronized cluster, possessing different frequencies and phases. Hence, $k$-solitary states comprise $k$ isolated elements \cite{k_Solitary}. Recently, the existence of solitary states has been demonstrated in a network of ensembles having attractive and repulsive interactions at the edge of synchrony~\cite{maistrenko} and partial synchrony~\cite{Teichmann2019}, inertial Kuramoto model~\cite{jaros}, oscillators with negative time-delayed feedback under external forcing~\cite{9_Second_Order}, identical populations of Stuart-Landau oscillators~\cite{premlatha}, FitzHugh-Nagumo neurons in the oscillatory regime~\cite{Soli_FN1} and neuronal oscillators and coupled chaotic maps in the presence of delayed links~\cite{Schulen2019}. The occurrence of solitary states can be observed in power grid networks in which individual grid-units gradually desynchronize during a partial or complete blackout \cite{Hellmann}. 

Furthermore, in real-world complex systems, a set of interacting units may have different types of interactions among them, with each type of interaction affecting functionality of other types. In such scenarios, the multiplex (multilayer) framework turns out to be an apt contender in representing different dynamical processes acting on the same set of units through different layers comprising different genres of links having different connectivity among the same set of interlinked nodes \cite{Ghosh2018, Domenico2017, Buldu2018}. Recently, the investigations pertaining to the emergence of chimera states and solitary states have been extended to multilayer networks subjected to a variety of dynamical models~\cite{Ghosh2016, Sawicki2019, Omelchenko2019, Pournaki2019, Majhi2017, Majhi2019, Mikhaylenko2019}.

The occurrence of chimera or solitary states have been designed through experimental setup comprising Huygens clock mechanical oscillators \cite{Kapitaniak2014}, coupled candle-flame oscillators via quenching and clustering \cite{Krishna2019}, modular networks of electrochemical oscillations \cite{Luis2019}, and locally and non-locally coupled Stuart-Landau oscillator circuits \cite{Gambuzza2020}.

Furthermore, machine learning techniques have been successfully being applied for prediction of system properties or emergent phenomena covering a broad areas of interdisciplinary research which ranges from non-linear dynamics, quantum physics, astrophysics to bio-medics \cite{Mitchell1997, Burkov2019,flach2020}. The field of complex systems and nonlinear dynamics has also witnessed a recent spurt in the use of machine learning techniques, particularly in characterization or identification of a variety of system properties or phenomena. For instance, the machine learning algorithms have been successfully implemented in community detection in networks \cite{Ghasemian2016}, finding fixed points attractors \cite{Rivkind2017}, spatiotemporal chaotic systems \cite{Jaideep2018}, detecting phase transition \cite{Ni2019}, prediction of chaotic systems \cite{Fan2020} and identification of chimera states \cite{Ganaei2020}.

In the present work, by employing machine learning techniques we predict the value of delay for engineering  chimera or solitary state for a given set of systems parameters. First off, we generate chimera state and solitary state in two altogether different network architectures, 2-D lattice and multiplex network. The presence of delay in neighboring connections of a node in a 2-D lattice structure gives rise to ripples of wave like chimera states, labelled as rippling chimera.
Whereas, for the occurrence of solitary states, inter-layer connections delays of a multiplex network is established which prohibits few individual node to fall within the synchronized clusters of identical nodes. 
Note that the presence of delays in either neighboring links in 2-D lattice or inter-links in multiplex network induces perturbations only in the dynamical evolution of the nodes and does not compromise with the structural symmetry of either network. 
Thereafter, for given data sets we employ multiple machine learning algorithms to train a model which is then used to predict the critical value of delay for yielding the chimera state and the intensity of the rippling chimera for a given choice of system's parameters. The K-nearest neighbours (KNN), support vector machine (SVM), and multi-layer perceptron neural network (MLP-NN) classifier are used utilizing the data generated from the two models. The analysis unveil that multi-layer perceptron neural network (MLP-NN) classifier is the best candidate in precisely predicting the critical delay values for engineering chimera and solitary states. Finally, we plot the entire phase space diagram using the trained machine learning model describing the parameter regimes having chimera and non-chimera states.

\section{Method and Technique}\label{method} 
This article considers two different coupled dynamics on network models to demonstrate the implementation of machine learning techniques for predicting the value of delays to design the solitary and the chimera states. Furthermore, using the trained machine learning model a more refined phase plots describing various dynamical states for the entire parameter region are plotted. In the following, first we discuss the coupled dynamics on  network models to demonstrate occurrence of chimera and solitary states by introducing delays in the coupling between pairs of oscillators in their respective network structures. Thereafter, we will describe the machine learning techniques used here to create a model for predicting delay values for engineering chimera and solitary states.\\

\subsection{Chimera states in 2-dimensional lattice}
\begin{figure}[t!]
    \centering
     \includegraphics[scale=0.33]{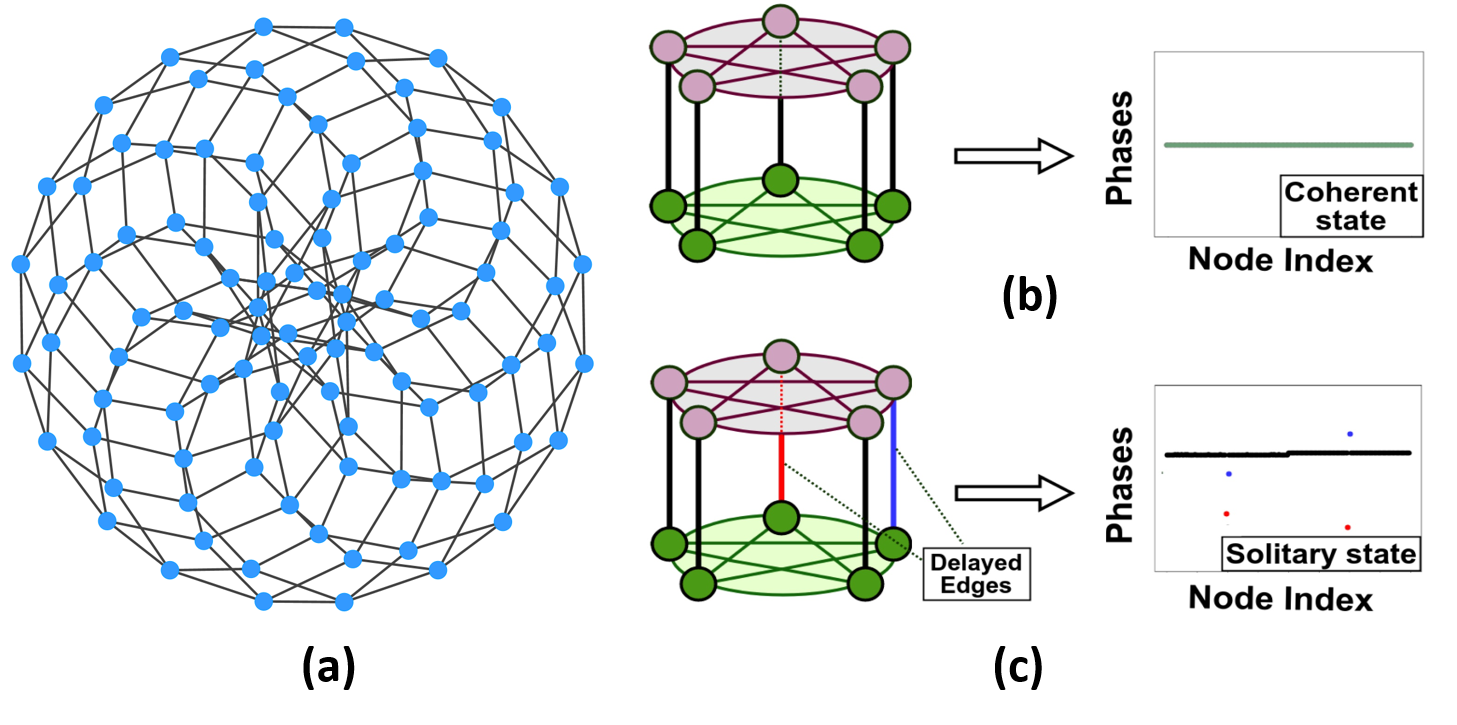}
    \caption{(a) Schematic diagram of $100$ identical nodes networked in a 2-D lattice formation assuming periodic boundary condition. (b) Schematic diagram of a multiplex network of two globally coupled networks having mirror inter-layer links. For a suitable choice of systems parameters, the state of multiplex network (b) demonstrating coherence for undelayed inter-layer links and (c) demonstrating solitary states in the presence of a few delayed interlayer links.}
    \label{bothNetwork}
\end{figure}

We consider $N$ nodes, each having $4$ nearest neighbors, arranged in a 2-dimensional lattice formation assuming periodic boundary condition (see Fig.~\ref{bothNetwork}(a)), with the local dynamics at each node governed by the Kuramoto oscillators. Such 2-D lattice exhibits coherence at large coupling strength. However, when a delay is introduced in each neighboring link of a randomly selected node, referred as delayed node, the delayed node and its neighboring then start exhibiting incoherence in the synchronous chunk, thus giving rise to chimera. The presence of delays in the coupling links for a node means that the information the node receives from its neighbors are delayed in time.

Thus the evolution of phase of an un-delayed and a delayed node $i$ is respectively given by
\begin{eqnarray}
\dot{\theta_{i}}=\omega+\mu\sum_{j=1}^{N}A_{ij}\sin{(\theta_j(t-\tau_{i})-\theta_i(t))}, \label{3}
\end{eqnarray}
where $\theta_i$ ($i{=}1,\ldots,N$) denotes the phase of $\textit{i}^{th}$ node, $\omega$ denotes the identical intrinsic frequency of the nodes, $\mu$ is the coupling strength and $\tau_{i}$ is the value of delay introduced in the links of $\textit{i}^{th}$ node. For delayed nodes $\tau_{i}\neq0$ and for non-delayed nodes $\tau_{i}=0$. $A_{ij}$ is the element of adjacency matrix of the network defined as
\begin{gather*}
A_{ij}=
\begin{cases}
1 & \text{if i and j are connected},\\
0 & \text{if, otherwise.}
\end{cases}
\end{gather*}

\subsection{Solitary states in multiplex network}
To demonstrate the occurrence of solitary states by setting up a discrete arrangement of delayed inter-layer links in a multiplex network, we begin with considering a multiplex network of two identical globally connected rings (of size $N$) whose nodes obey the dynamics of Kuramoto oscillators.
Figs.~\ref{bothNetwork}(b) and \ref{bothNetwork}(c) illustrate a schematic representation of a multiplex network in the absence and the presence of delayed inter-layer links. In the absence of delay, both the layers of the multiplex network are in coherent state at sufficiently large coupling strength. However, as the delay is established in one of its inter-layer links, two end nodes of the delayed link then get dislodged from their coherent state.

To fabricate the delayed environment in the system, we take into account delayed couplings at a number of arbitrarily chosen but discretely located inter-layer links in the multiplex network, referred to as inter-layer delays for the sake of convenience. Hence, time update of the dynamics of Kuramoto oscillators in the multiplexed identical layers $1$ and $2$ under the delayed setting is governed by: 
\begin{align}
	\begin{split}
	\dot{\theta_{i}^{1}} =  \omega + \mu_1 \sum_{j=1}^{N} {}& \sin(\theta_{j}^{1} - \theta_{i}^{1}) + \\
	& \sigma^{12} \sin(\theta_{i}^{2}(t-\tau_i) - \theta_{i}^{1}(t)), \label{eq:eqn1}
	\end{split}\\
	\begin{split}
	\dot{\theta_{i}^{2}} = \omega + \mu_2 \sum_{j=1}^{N} {}& \sin(\theta_{j}^{2} - \theta_{i}^{2}) + \\
	& \sigma^{21} \sin(\theta_{i}^{1}(t-\tau_i) - \theta_{i}^{2}(t)), \label{eq:eqn2}
	\end{split}
\end{align}
where $\omega$ is the identical intrinsic frequency, $\theta_i$ is the phase of $i^{th}$ ($i{=}1,...,N$) node, $\mu$ is the intra-layer coupling strength of a layer and $\sigma^{12}=\sigma^{21}=\sigma$ is the inter-layer coupling strength representing the impact of dynamics of one layer on the other. $\tau_i$ is an element of a delay-vector $\tau$ (of length $N$), which contains particulars about the position and the amplitude of delayed inter-layer links and is directed along either $\sigma^{12}$ or $\sigma^{21}$. We assume a fraction $N_{\tau}$ (significantly smaller than $N$) of arbitrarily picked discrete locations in $\tau$, which contains time-delays ($\tau_i\neq0$) either drawn from a uniform random distribution or of identical amplitude. The remaining fraction ($N-N_{\tau}$) of $\tau$ contains no delay, i.e., $\tau_i=0$. Therefore, the number $N_{\tau}$ determines the number of coveted solitary states, i.e., the system can have 1-solitary state, 2-solitary states, or maximum possible $N_{\tau}$-solitary states in a layer. Also, the fraction of heterogeneous (identical)  ($\tau_i\neq0$) delays in $\tau$ would give rise to solitary points with unequal (equal) phase displacement from the synchronous cluster.

\subsection{Machine learning techniques}
In this paper, three different supervised machine learning algorithms are employed to predict the precise value of delay to engineer solitary and chimera states for a given set of network parameters. These machine learning algorithms are K-nearest neighbours (KNN) classifier, support vector machine (SVM) classifier and multi-layer perceptron neural network (MLP-NN) classifier.

KNN classifier is a non-parametric classification algorithm, which has been proven to be effective in numerous cases. If we represent our data in a vector space, each point in this vector space can be classified based on the classes of $k$ nearest neighbors of the data point. The $k$ nearest neighbors are selected based on a distance parameter. Most commonly, the euclidean distance is used to determine the $k$ nearest neighbors. Therefore, KNN divides our data's vector space into different regions corresponding to different classes. The parameter $k$ plays a very important role in deciding how well KNN will perform while dividing the vector space into different regions and classifying the points in that vector space~\cite{murphy2012machine}.

SVM classifier is a supervised machine learning model, which performs by estimating the most appropriate hyperplane that can separate our training data into two different distinct classes. The hyperplane estimation is achieved by maximizing the distance between the nearest training data point and the proposed hyperplane. This distance is also called margin. Simple SVM can only produce linear hyperplanes. One can use kernels to estimate nonlinear hyperplanes. A kernel functions by transforming our training data from a lower-dimensional space into a higher-dimensional space and estimating a linear hyperplane in that higher-dimensional space. When the higher-dimensional hyperplane is transformed back to the lower-dimension, we get a nonlinear hyperplane that can classify each point of our data's vector space into different classes~\cite{murphy2012machine}.

MLP-NN classifier functions by creating an artificial neural network consisting of many different layers of nodes. There exist three types of layers in a neural network, the input layer, hidden layer and the output layer. One can have any number of hidden layers, and each hidden layer can have any number of nodes. The neural network takes the input data and tries to estimate the weights of each link between the nodes of the network. A neural network can be called a trained model if the algorithm can successfully estimate the weights of the links such that the model can categorize our data into their correct classes~\cite{murphy2012machine}.

\section{Results}
First off, we numerically demonstrate the occurrence of chimera and solitary states in the models discussed in section \ref{method}. Thereafter, we make predictions for the precise value of critical delay required for the engineering chimera states and solitary states by employing machine learning classifiers.
Here, we numerically demonstrate the occurrence of chimera and solitary states in two distinct network structures.

\subsection{Engineering chimera states}\label{simulation}
\begin{figure}[t!]
    \centering
    \includegraphics[scale=0.33]{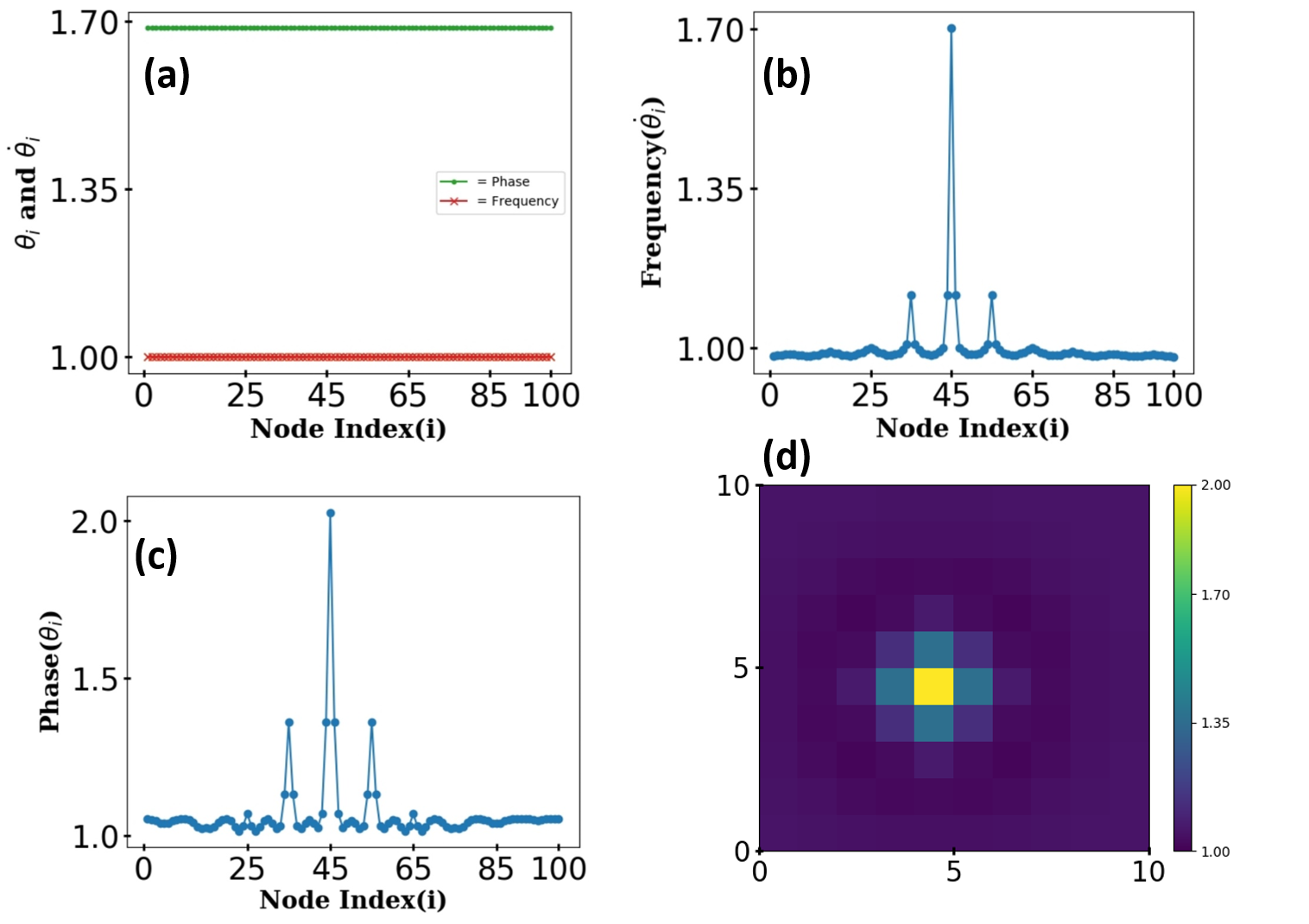}
    \caption{Phase and frequency snapshot of 2-D lattice of $N=100$ nodes and $\mu=1$. (a) Phase and frequency snapshot of the nodes in the absence of delay ($\tau=0$). (b) A frequency snapshot and (c) a phase snapshot of the nodes in the presence of delay ($\tau=10$) in all the links to \nth{45} node. (d) A heatmap representation of the phases in (c). (b), (c) and (d) demonstrate rippling chimera states.}
    \label{chimera}
\end{figure}

To engineer chimera states, the intrinsic frequency of the $N=100$ nodes are considered to be the same, i.e., $\omega=1$ and their initial phases are assigned randomly in the interval $[0, 2\pi)$. We begin with an un-delayed but synchronous 2-D lattice obtained for coupling constant $\mu=1$ as shown in Fig.~\ref{chimera}(a). Starting from a set of initial random phases, after sufficiently high intralayer coupling strength, all the oscillators settle into the steady phase with constant frequency ($\omega=1$).
However, when a delay is instituted in all the neighboring links to a node(say 45th node) of the lattice, this produces the perturbation in the neighboring links and hence giving rise to chimera states (see Figs.~\ref{chimera}(b) and \ref{chimera}(c)). Such emergent chimera pattern resembles to the ripples on the surface of water, originating from the delayed ($i^{th}$) node, hence is termed as rippling chimera states.

When the lattice evolves in the presence of delay at neighboring links to a node (say \nth{45} node), as expected the phase and the frequency of the delayed node (\nth{45}) and its neighboring nodes (for example: \nth{35} and \nth{46}) get desynchronized from their respective synchronous clusters (rest of the nodes). Nevertheless, the rest of the nodes remain synchronous and their frequency still closely follows the intrinsic frequency. Moreover, a gradual steep fall in the amplitude of both the phase and frequency starting off the delayed node through neighboring nodes until the rest synchronous chunk is quite apparent from Figs.~\ref{chimera}(b) and \ref{chimera}(c), mimicking the ripples on the surface of synchronous cluster. This phenomena is also reflected from Fig.~\ref{chimera}(d), the heatmap representation of the phases of the nodes for a delay present in the neighboring links of \nth{45} node. The desynchronized delayed node and its neighboring nodes are referred here as drifting oscillators.
Thus, the inclusion of delayed links to a node give rise to the rippling chimera states, whereas the delayed node and its neighbors form the incoherent regime and the rest of the nodes remain part of the synchronous regime.

\subsection{Engineering Solitary States}
\begin{figure}[t!]
	\begin{center}
		\includegraphics[scale=0.05]{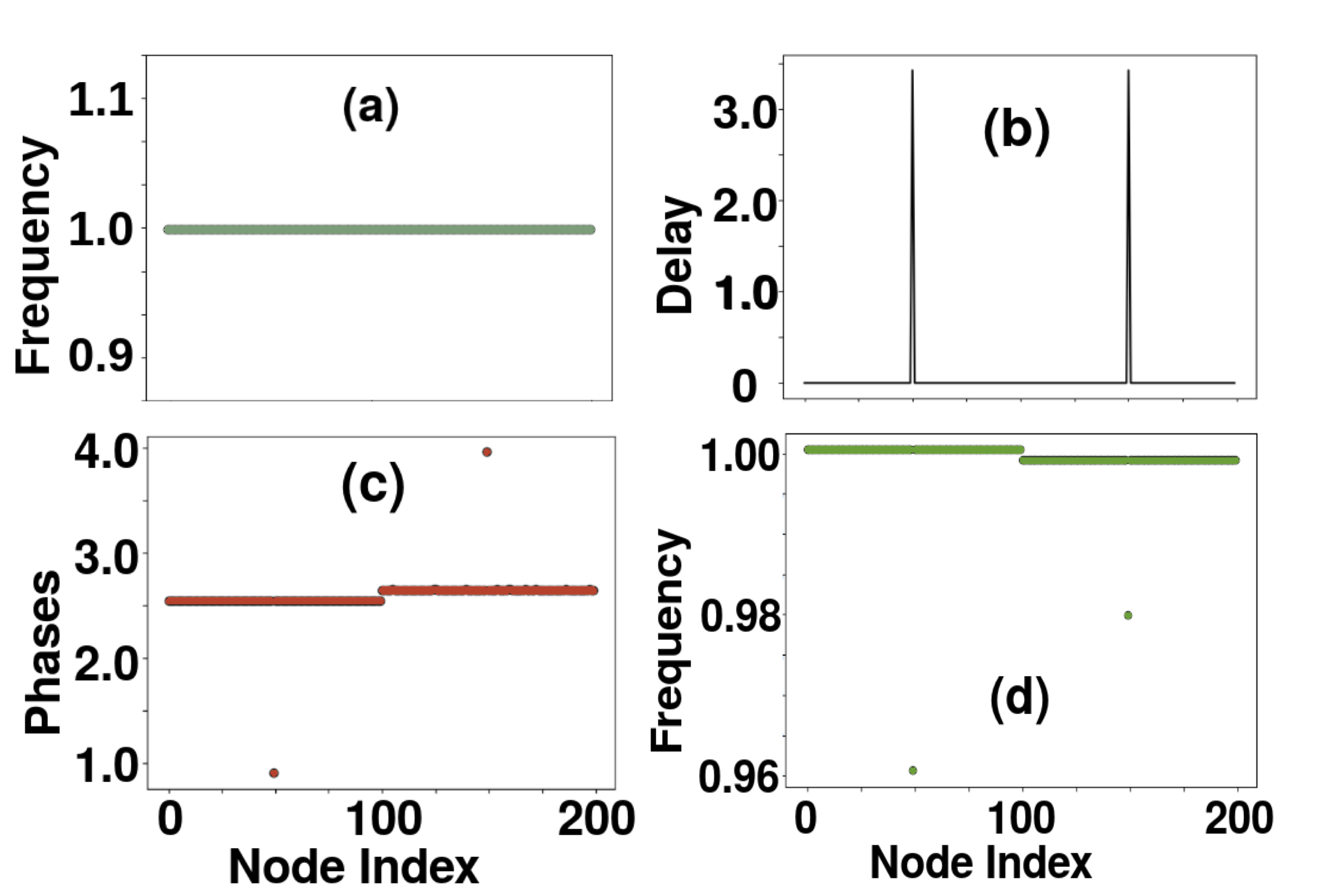}\\
	\end{center}
	\caption{Snapshots of the layers of the multiplex network displaying (a) a flat frequency profile with undelayed inter-layer links (b) introduced delay profile at inter-layer links (c) Phase and (d) frequency profile for the solitary state with introduced delayed interlayer links. System parameters are $\mu_1 {=} 0.5$, $\mu_2 {=} 3$ and $\sigma {=} 1$.}
	\label{fig:figure2}
\end{figure}

\begin{figure}[t!]
	\begin{center}
		\includegraphics[scale=0.025]{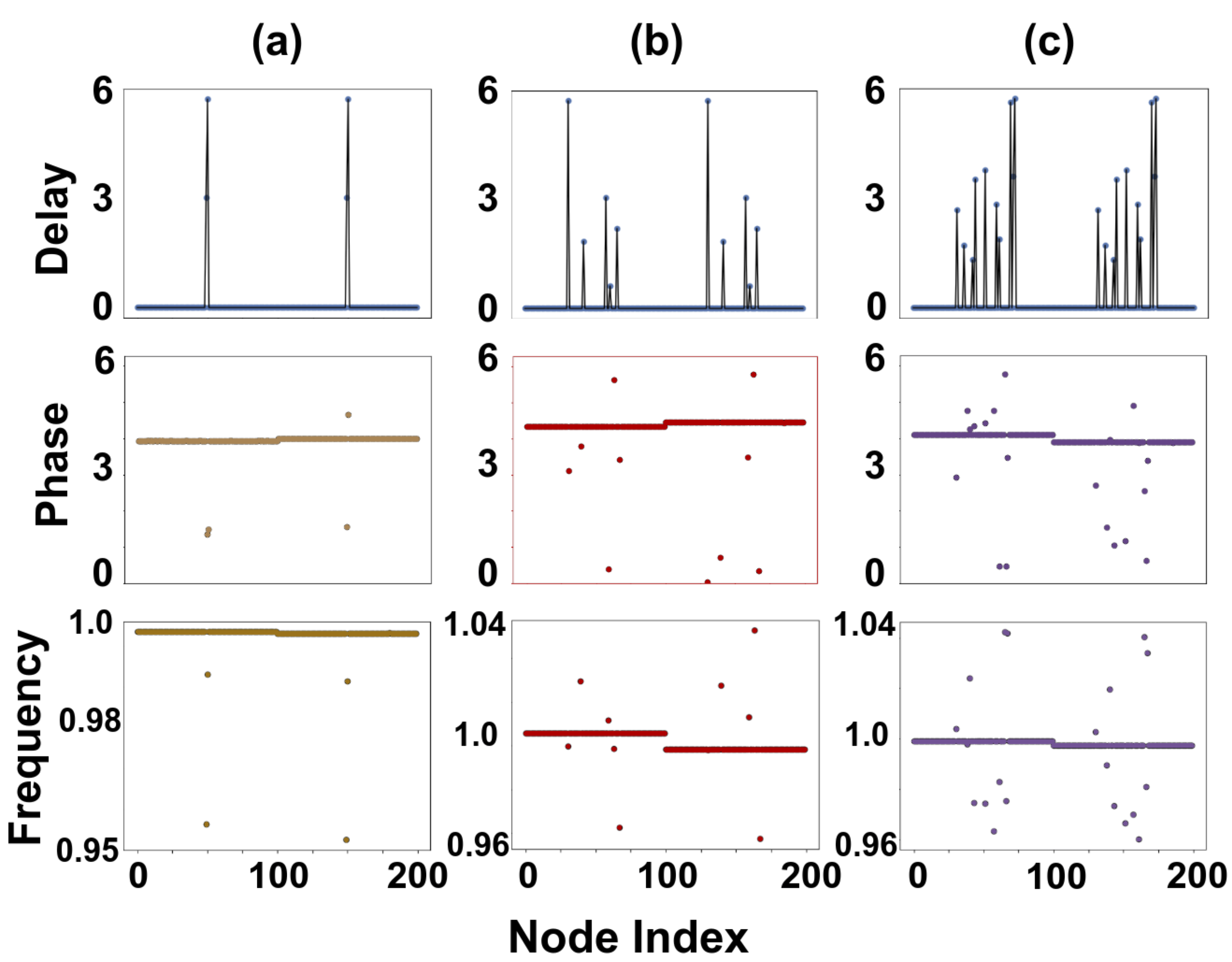}\\
	\end{center}
	\caption{Snapshots of the layers of the multiplex network with (top row) different inter-layer delay profile, resulting in (mid row) Phase; (bottom row) frequency profile for (a) 2- (b) 5- (c) 10- solitary states with the introduced inter-layer delays (as depicted in top row), respectively.}
	\label{fig:figure3}
\end{figure}

Solitary states are spatiotemporal patterns obtained from the dislodgement of a few nodes from the main synchronous cluster, which possess frequencies different than that of the synchronous cluster. To exhibit the emergence of solitary states in the multiplex network with the aid of inter-layer delays, we select initial phases of the nodes drawn randomly from the interval $\theta_i^{1,2}\in[0, 2\pi)$. We start off with an un-delayed but synchronized multiplexed rings, each of $100$ nodes, which is obtained for intralayer coupling constants $\mu_1 = 0.5$, $\mu_2 = 3$ and interlayer coupling constant $\sigma = 1$ as shown in Fig.\ref{fig:figure2}(a). 
Now the presence of delay in one of the interlayer links (with end nodes $i,i=50, N{+}50$) exhibits dislodgement of the phases (frequencies) of the interconnected nodes from their respective phase (frequency) synchronized clusters (see Figs.~\ref{fig:figure2}(c) and ~\ref{fig:figure2}(d)) resulting in two 2-solitary states, one for each layer. Note that the choice of $\mu_1=\mu_2$ (one yielding synchronous clusters) can also result in splitting off phases from the main synchronous clusters; however, this does not induce dislodgement in frequencies of the same nodes, hence can not be delineated as solitary states. A mismatch in intra-layer coupling strength $\mu_1$ and $\mu_2$ ensures splitting off the frequencies along with the phases of the end nodes of inter-layer delays. In similar fashion, $2N_{\tau}$ solitary states are accrued from the presence of $N_{\tau}$ delayed inter-layer links as shown in Fig.\ref{fig:figure3}. The column panels (a), (b) and (c) in Fig.\ref{fig:figure3} corresponding to $N_{\tau}=2, 5\ \mbox{and}\ 10$ exhibit 4, 10 and 20-solitary states.

Delay is integral to our scheme to get solitary states. The phase difference between the dislodged nodes and the bulk of synchronous nodes can be different; however, the corresponding frequency mismatch remains almost the same for any value of delay for a set of structural parameters $\mu_1, \mu_2$ and $\sigma$. The presence of delay in an inter-layer link makes the dynamics of the nodes at its two ends either slower or faster than the rest bulk of synchronized nodes. Therefore, the employed scheme allows us to settle on the appropriate values for the delay and the inter-layer coupling strength($\sigma$), which can substantially change the frequencies of the end nodes of the inter-layer delays than those of the rest of the nodes, yielding pronounced solitary states.


Note that besides generating tailored solitary states, the employed scheme can generate chimera states as well when the fraction of delays are installed in a string of inter-layer links instead of globally spread ones. 

\subsection{Implementation of Machine learning  techniques}
In this section, three different supervised machine learning algorithms trained on the input data obtained from simulations in \ref{simulation} are used for the model-free prediction of factors determining or controlling the intensity of chimera and solitary states emergent in different network structures.

\subsubsection{Predicting intensity of rippling chimera and critical delay using machine learning}
\begin{figure}[t!]
    \centering
    \includegraphics[scale=0.35]{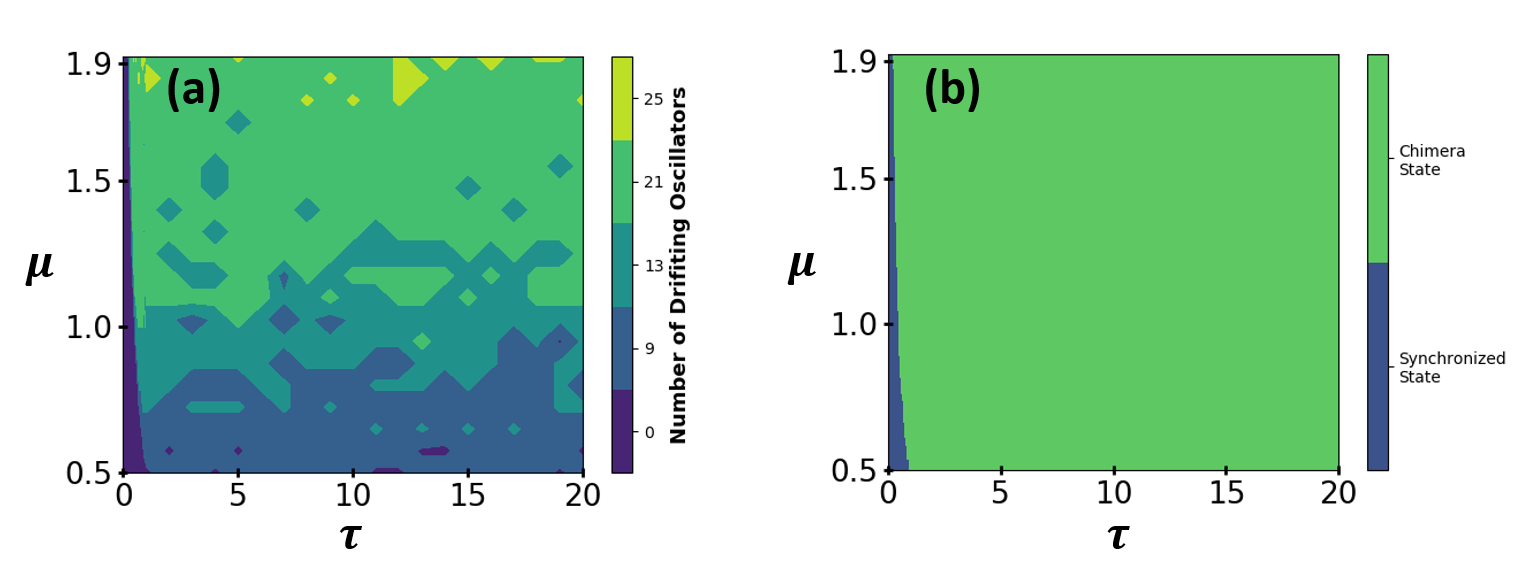}
    \caption{(a) Phase diagram for 2D lattice in two parameter space of $\tau$ and $\mu$. This phase diagram is plotted using the data which is directly obtained from the simulations. (b) This is a filtered version of (a). All the regions in (a) with positive values for "number of drifting oscillators" are merged here to form green region, which represents chimera state. The number of drifting oscillators in blue region is zero, which represents a synchronized state.}
    \label{phasePlot}
\end{figure}

Here, we make the prediction for the intensity of chimera states using machine learning classifiers discussed in section~\ref{method}. To generate the data used in training the machine learning models, the network is allowed to evolve in time
(using RK-4 algorithm with time-steps $\Delta$t=0.01) for different values of coupling constant and delay. The coupling constant is varied from 0.5 to 2 with interval of 0.075. The delay is varied from 0 to 1 with interval of 0.05 and from 1 to 20 with interval of 1. Total $800$ simulations are carried out and the number of drifting oscillators is recorded at the end of each simulation. Phase diagram is then plotted using the raw data obtained from the simulations (Fig.\ref{phasePlot}(a)).
Fig.\ref{phasePlot}(a) unveils that the data contains a lot of noise, which arises due to the inaccuracies in the numerical simulations. From the inspection of data, only one boundary can be drawn with certainty as shown in Fig.\ref{phasePlot}(b). This diagram provides a parameter space for which synchronized and chimera regimes are distinguishable, however it lacks in capturing some useful information that our data contains. For instance, the exact number of drifting oscillators or the intensity of chimera state can not be discovered by the inspection of this diagram.

To construct a more detailed phase diagram, machine learning techniques are used.
The data is tabulated in three columns where first, second and third column contain the value of $\mu$, $\tau$ and number of drifting oscillators corresponding to the pair of delay and coupling constant, respectively. The data structure looks like Table~\ref{ripple data table}. The data is randomly split into the training and the testing set in the ratio of 4:1. Our task here is to train a machine learning model to predict the number of drifting oscillators for an input pair of $\mu$ and $\tau$.
The number of drifting oscillators is calculated for each pair of $\mu$ and $\tau$ at the end of numerical simulation as shown in Table~\ref{ripple data table}. The number of drifting oscillators is determined from the number of points in the node index vs frequency plot(ex. Fig(\ref{chimera}b)), which are far apart from the natural frequency of the oscillators.

Machine learning algorithms have hyperparameters which determine how well a trained machine learning classifier will perform on a given data. In order to get the best possible predictive model which an algorithm is capable of generating, one needs to find the optimal hyperparameters for that algorithm, which depend on the given dataset. One by one, we provide the details of hyperparameters for the three algorithms put into practice.

\begin{figure}[t!]
    \centering
    \includegraphics[scale=0.45]{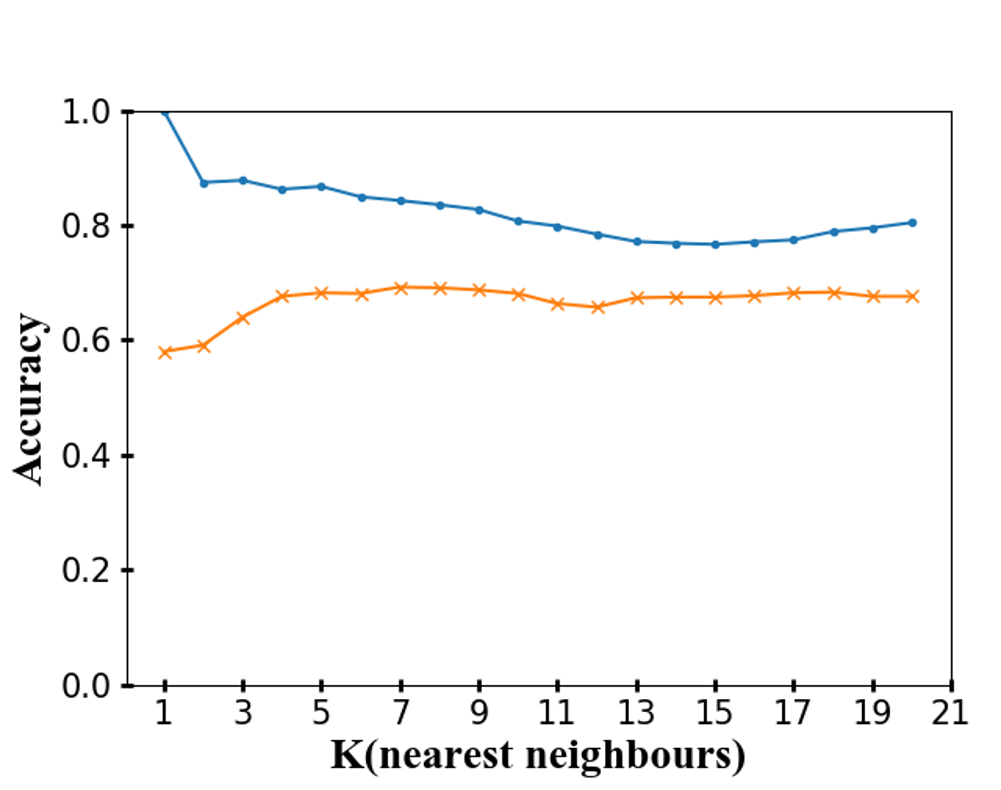}
    \caption{Validation curve for KNN obtained using 5 fold cross validation of data. The parameter $K$ is varied from 1 to 20. The value of $K$ at which a KNN model yields high training accuracy as well as high validation accuracy is the optimum value of $K$ for the dataset. $K=1$ leads to overfitting as in that case the training accuracy is 1 but the validation accuracy is very low. The validation curve suggests that the best possible KNN model that one can obtain for our dataset is for $K=5$. Blue and orange line represent training set accuracy and validation set accuracy, respectively.}
    \label{validationCurveKnn}
\end{figure}

(a) {\em KNN}: The validation curve is plotted to find the optimal hyperparameter $K$ for KNN as shown in Fig~\ref{validationCurveKnn}. The optimal hyperparameters for KNN is given in Table~\ref{latticeParameters}.


(b) {\em SVM}: SVM has three hyperparameters which are kernel, regularization parameter and gamma. To find the optimal hyperparameters for SVM, the grid search analysis is performed. The optimal hyperparameters for SVM is given in Table~\ref{latticeParameters}.


(c) {\em MLP-NN}: All the optimal hyperparameters and other information about the MLP-NN is given in Table~\ref{latticeParameters}. 

Next 1000, 1000 and 100 models are generated for KNN, SVM and MLP-NN, respectively, by choosing different training sets at random for each iteration. The final prediction for the number of drifting oscillators using KNN, SVM and MLP-NN classifiers is obtained by aggregating the results of these 1000 KNN, 1000 SVM and 100 MLP-NN models.

\begin{figure}[t!]
    \centering
    \includegraphics[scale=0.35]{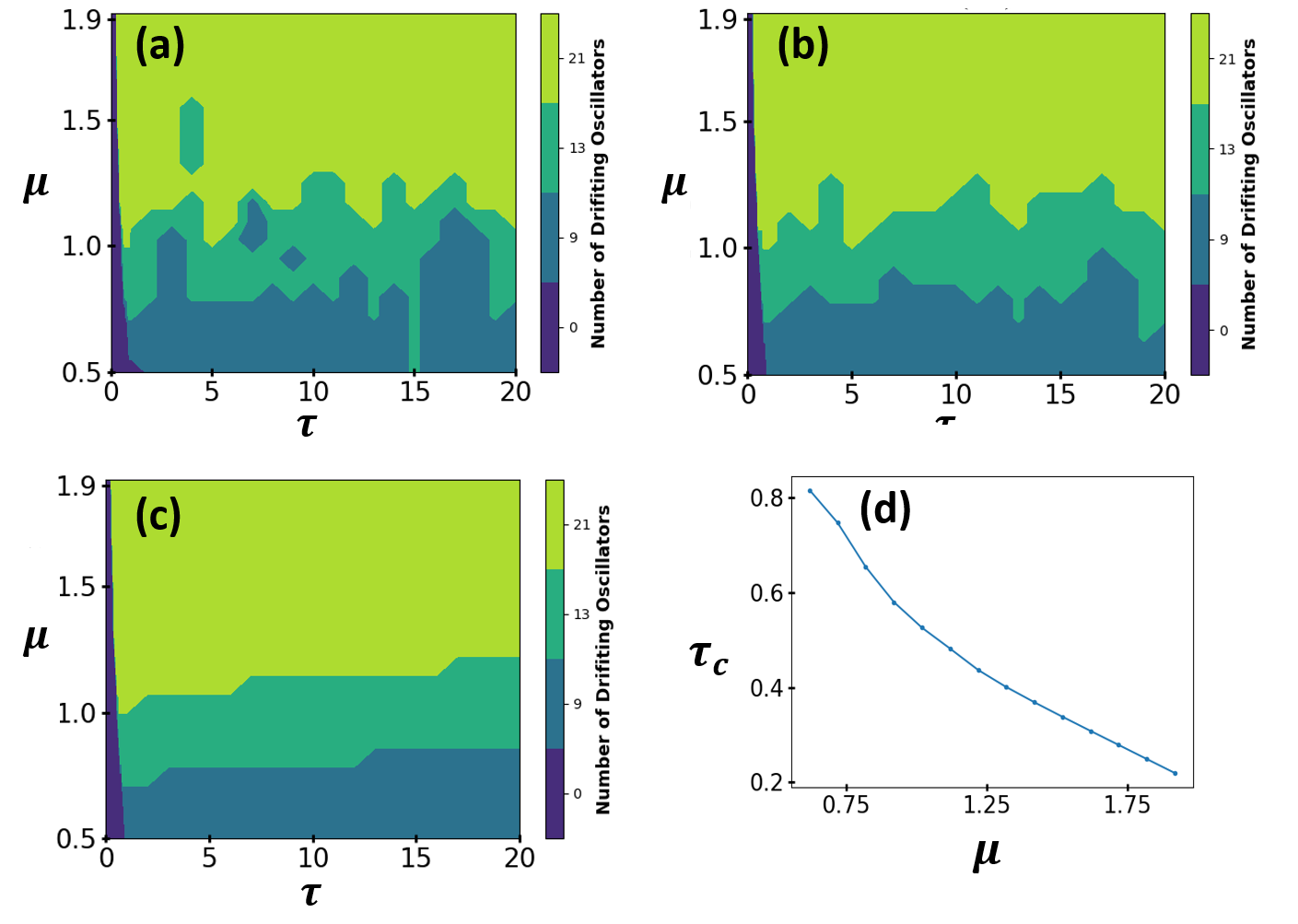}
    \caption{Phase diagrams of 2D lattice in $\tau$ and $\mu$ parameters space employing (a) KNN, (b) SVM and (c) MLP-NN algorithm. Here, the region boundaries are determined using a trained machine learning model. These are filled contour plots where each colour represents a different value for "number of drifting oscillators" found in the engineered chimera state. (a) Using KNN algorithm: the value of parameter K=5 for training this model is obtained using validation curve analysis (see Fig.~\ref{validationCurveKnn}). (b) Using SVM algorithm: RBF kernel are used while training this SVM model. The value of parameters C=10 and gamma=0.5 is obtained used grid search analysis. (c) Using MLP-NN algorithm: ReLU activation function is used while training this MLP-NN model. Apart from the input and output layer, the artificial neural network used in this algorithm contains 2 hidden layers with 30 nodes each. (d) The behavior of critical delay as a function of $\mu$ for a 2D lattice network. The values of critical delay in this plot are calculated using a trained MLP-NN machine learning model.}
    \label{MLCombined}
\end{figure}

Training accuracy measures how well an algorithm learns from the training data whereas testing data measures how well an algorithm can train a model which can classify a data which it has never seen before. Higher value of testing accuracy is more desirable than higher value of training accuracy. High training accuracy but low testing accuracy can mean that the algorithm is over-fitting our data. Table \ref{accuracy table} shows that out of the three algorithms, neural network is the best at classifying any unknown data. The lower value of training accuracy for MLP-NN is due to the fact that MLP-NN is able to identify the noise present in the training data (Fig~\ref{phasePlot}a). It classifies the noisy data-points into their correct classes (Fig~\ref{MLCombined}c), which lowers it's training accuracy as compared to KNN (Fig~\ref{MLCombined}a) and SVM (Fig~\ref{MLCombined}b) which are not good at classifying the noisy data-points into correct classes.
The phase diagrams obtained using each algorithm (Figs.~\ref{MLCombined}a,~\ref{MLCombined}b,~\ref{MLCombined}c) shows that out of the three algorithms, neural network is significantly better at segregating different regions in a phase space ($\tau$-$\mu$) corresponding to different intensities of chimera.
Figs.~\ref{confusionMatrix}a,~\ref{confusionMatrix}b and~\ref{confusionMatrix}c show the confusion matrix for KNN, SVM and MLP-NN classifiers, respectively. Tables~\ref{sensitivity} and~\ref{specificity} show the sensitivity and specificity of each algorithm, respectively. Comparisons of sensitivity and specificity for each algorithm confirm that MLP-NN is the best one out of the three algorithms in identifying the noise present in the training data and classifying the noisy data-points into their correct classes.
Therefore, MLP-NN algorithm stands out in predicting the intensity of chimera state in a system.

For a value of coupling constant, the minimum value of delay which transitions the network from a synchronized to chimera state is known as critical delay of the network corresponding to that value of coupling constant. 
The trained MLP-NN machine learning model was used to predict the exact values of critical delay for a set of coupling constant values(Fig.~\ref{MLCombined}d).
To predict the exact value of critical delay corresponding to a coupling constant, $\mu$ is kept fixed and $\tau$ is increased from $\tau=0$ in the steps of $0.001$ and the final collective behaviour of the network for each pair of $\mu$ and $\tau$ is predicted using the MLP-NN model. The smallest value of $\tau$ for which the MLP-NN model predicted the final collective state to be a chimera, is the critical delay corresponding to the given value of $\mu$.
Using this technique, critical delay corresponding to any value of coupling constant can be found. The advantage of using a ML model to find critical delay is that it is very fast as we don't have to run any simulation once the machine learning model is trained, to predict those values of critical delays.

\begin{table}
\caption{Data structure for Rippling data: There are 800 rows in total in this table.}
\begin{center}
\scalebox{0.9}{
\begin{tabular}{|c|c|c|}
        \hline
        Coupling constant & Delay & No. of drifting oscillators \\
        \hline
        : & : & : \\
        : & : & : \\
        0.875 & 0.5 & 0 \\
        0.875 & 0.6 & 9 \\
        0.875 & 0.7 & 13 \\
        : & : & : \\
        1.1 & 1 & 21 \\
        1.1 & 2 & 21 \\
        : & : & : \\
        : & : & : \\
        \hline
        \end{tabular}}
\end{center}
\label{ripple data table}
\end{table}

\begin{table}
\caption{Accuracy of different algorithms}
\begin{center}
\scalebox{0.85}{
\begin{tabular}{|l|l|l|}
\hline
Algorithm      & Training accuracy & Testing accuracy \\ \hline
KNN            & 85.449            & 77.66            \\ \hline
SVM            & 85.455            & 80.821           \\ \hline
Neural Network & 83.012            & 82.725           \\ \hline
\end{tabular}
        }
\end{center}
\label{accuracy table}
\end{table}

\begin{table}
\caption{Sensitivity of different algorithms for 2D lattice network}
\begin{center}
\scalebox{0.68}{
\begin{tabular}{|l|l|l|l|l|l|l|l|l|}
\hline
               & \multicolumn{8}{c|}{Sensitivity}                                               \\ \hline
Algorithm      & Zero  & One & Five & Nine  & Thirteen & Twenty One & Twenty Five & Twenty Nine \\ \hline
Knn            & 1     & 0   & 0    & 0.817 & 0.676    & 0.949      & 0.166       & 0           \\ \hline
SVM            & 0.995 & 0   & 0    & 0.796 & 0.669    & 0.955      & 0.166       & 0           \\ \hline
Neural Network & 0.984 & 0   & 0    & 0.72  & 0.633    & 0.964      & 0           & 0           \\ \hline
\end{tabular}
        }
\end{center}
\label{sensitivity}
\end{table}

\begin{table}
\caption{Specificity of different algorithms for 2D lattice network}
\begin{center}
\scalebox{0.68}{
\begin{tabular}{|l|l|l|l|l|l|l|l|l|}
\hline
               & \multicolumn{8}{c|}{Specificity}                                                 \\ \hline
Algorithm      & Zero  & One   & Five & Nine  & Thirteen & Twenty One & Twenty Five & Twenty Nine \\ \hline
Knn            & 0.98  & 0.999 & 1    & 0.965 & 0.953    & 0.902      & 1           & 1           \\ \hline
SVM            & 0.975 & 1     & 1    & 0.967 & 0.958    & 0.892      & 1           & 1           \\ \hline
Neural Network & 0.969 & 1     & 1    & 0.966 & 0.953    & 0.879      & 1           & 1           \\ \hline
\end{tabular}
        }
\end{center}
\label{specificity}
\end{table}

\begin{table}[]
\caption{Parameters for Machine learning models trained using dataset of 2D lattice network}
\begin{center}
\scalebox{0.58}{
\begin{tabular}{|c|c|c|c|c|c|}
\hline
\multicolumn{2}{|c|}{KNN}                                     & \multicolumn{2}{c|}{SVM}                                         & \multicolumn{2}{c|}{MLP-NN}                      \\ \hline
Description                      & Value                      & Description                               & Value                & Description                          & Value     \\ \hline
\multirow{3}{*}{K}               & \multirow{3}{*}{5}         & \multirow{3}{*}{Regularization Parameter} & \multirow{3}{*}{10}  & Total number of layers               & 4         \\ \cline{5-6} 
                                 &                            &                                           &                      & Number of hidden layers              & 2         \\ \cline{5-6} 
                                 &                            &                                           &                      & Number of nodes in each hidden layer & 30        \\ \hline
\multirow{3}{*}{Weight Function} & \multirow{3}{*}{Uniform}   & \multirow{3}{*}{Kernel}                   & \multirow{3}{*}{EBF} & Number of nodes in the output layer  & 8         \\ \cline{5-6} 
                                 &                            &                                           &                      & Optimizer                            & Adam      \\ \cline{5-6} 
                                 &                            &                                           &                      & Learning Rate                        & $10^{-3}$ \\ \hline
\multirow{4}{*}{Distance Metric} & \multirow{4}{*}{Euclidean} & \multirow{4}{*}{Gamma}                    & \multirow{4}{*}{0.5} & L2 Penalty                           & $10^{-4}$ \\ \cline{5-6} 
                                 &                            &                                           &                      & Activation                           & ReLU      \\ \cline{5-6} 
                                 &                            &                                           &                      & Batch Size                           & 200       \\ \cline{5-6} 
                                 &                            &                                           &                      & Epochs                               & 200       \\ \hline
\end{tabular}
        }
\end{center}
\label{latticeParameters}
\end{table}

\subsubsection{Predicting value of critical delay for emergence of solitary state}
Here, we precisely forecast the value of critical delay required to delineate solitary states using the machine learning algorithms. First, the coupled dynamics Eq.~(\ref{eq:eqn1}) and Eq.~(\ref{eq:eqn2}) are allowed to evolve for different values of interlayer coupling strength($\sigma$) and intra-layer coupling strength of layer 1($\mu_1$). $\mu_2=3$ is kept fixed in all the analysis done in this section. In total 1600 such simulations are performed and the frequency difference between the delayed nodes and rest of the synchronized nodes is recorded at the end of each simulation. This is performed for 5 different values of delay, i.e, 0, 0.5, 1, 2, 4. Therefore, the total number of simulations performed are 8000. For a given pair of values of inter-layer and intra-layer coupling, the system can achieve solitary state if a necessary amount of delay is applied to the system. It is rather difficult and computationally demanding to find the exact value of delay at which the system transits from the synchronized to the solitary state for given values of inter-layer and intra-layer coupling strength.

For that matter, we use machine learning techniques to find the precise value of the critical delay for a given set of values of inter-layer and intra-layer coupling strength. 

The data that is used to train our model is tabulated in four columns where first, second, third and fourth column contain the values of interlayer coupling constant($\sigma$), intra-layer coupling constant($\mu_1$), delay($\tau$) and 0 (1) for the synchronized state (solitary state), respectively.
A system is solitary or not is decided by looking at the frequency difference between the excited (delayed) node and a synchronized node. If the frequency difference is more than threshold value of 0.01 the system's state is then delineated as solitary state. The data structure looks like Table~\ref{solitaryTable}.

In order to predict the values of critical delay corresponding to any given set of values for inter-layer and intra-layer coupling strength, one trains a machine learning model to predict the final state (synchronized (0) or solitary (1)) of the multilayer network after feeding the input values of $\mu_1$, $\sigma$ and $\tau$. The value of delay is increased by $0.001$ in each iteration while keeping the values of $\mu_1$ and $\sigma$ fixed. The prediction of MLP-NN model for each combination of $\mu_1$, $\sigma$ and $\tau$ is then recorded for each iteration. The lowest value of delay is recorded for which the network makes transition from a synchronized state to solitary state. This way the value of critical delay is obtained for a given pair of $\mu_1$ and $\sigma$.

\begin{figure}[t!]
    \centering
    \includegraphics[scale=0.35]{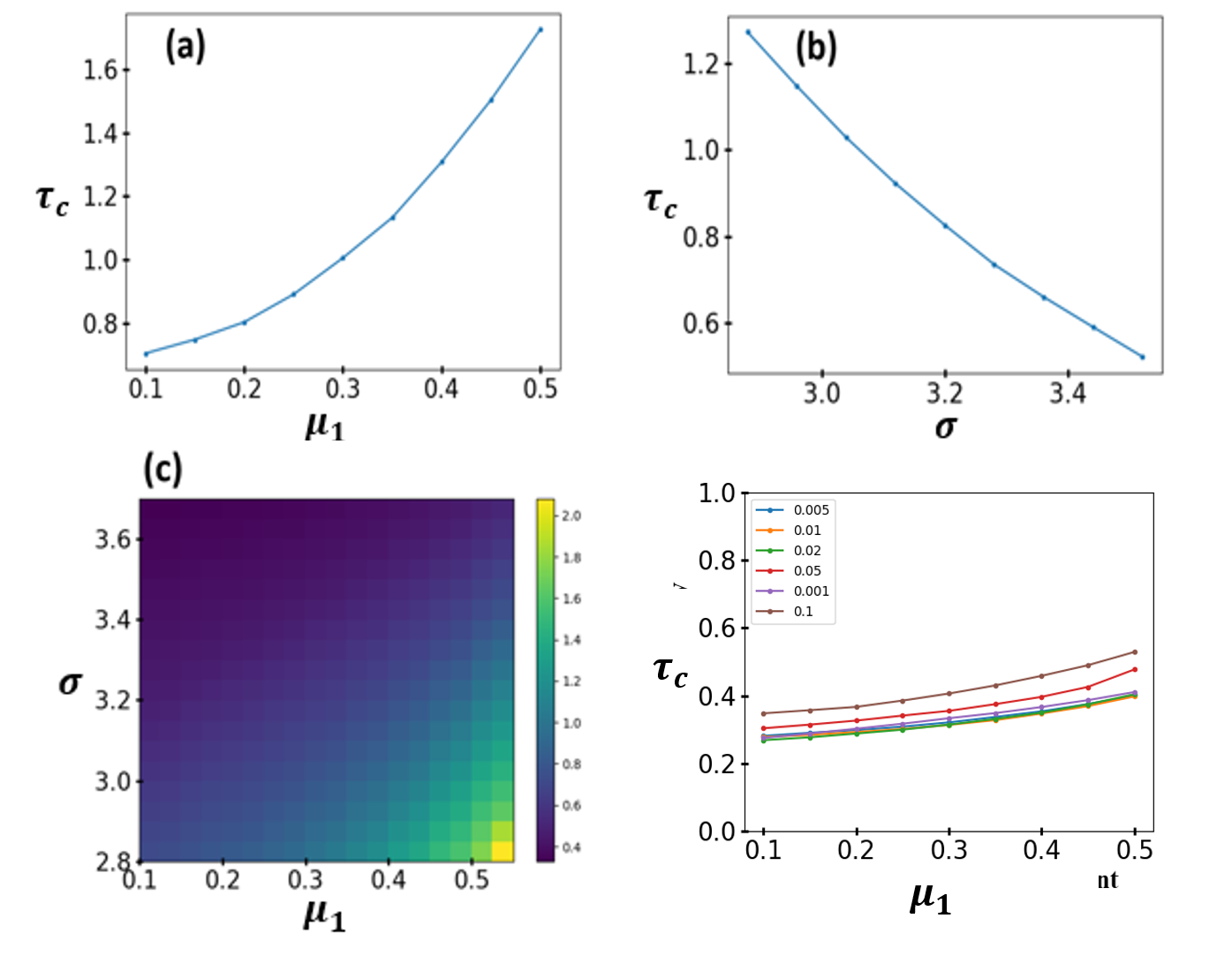}
    \caption{The behavior of critical delay as a function of (a) $\mu_1$ and (b) $\sigma$ for a multiplex network with $\sigma=2.81$ and $\mu_1=0.43$, respectively. (c) exhibits a heatmap in $\mu_1$ and $\sigma$ parameters space for the multiplex network. The colorbar represents the value of critical delay. The values of critical delay in (a), (b) and (c) are calculated using a trained MLP-NN machine learning model. (d) The critical delay as a function of $\mu_1$ for the multiplex network with $\sigma=3.82$. Each line corresponds to a different value of threshold for frequency difference between the desynchronized node and the synchronized nodes, to determine if a state is solitary or not. For a value of threshold between 0.001 and 0.02, the predicted value of the critical delay using machine learning model does not see any significant change. Once the value of threshold exceeds 0.02, the machine learning model starts yielding bad predictions.}
    \label{criticalDelay}
\end{figure}

\begin{figure}[t!]
    \centering
    \includegraphics[scale=0.38]{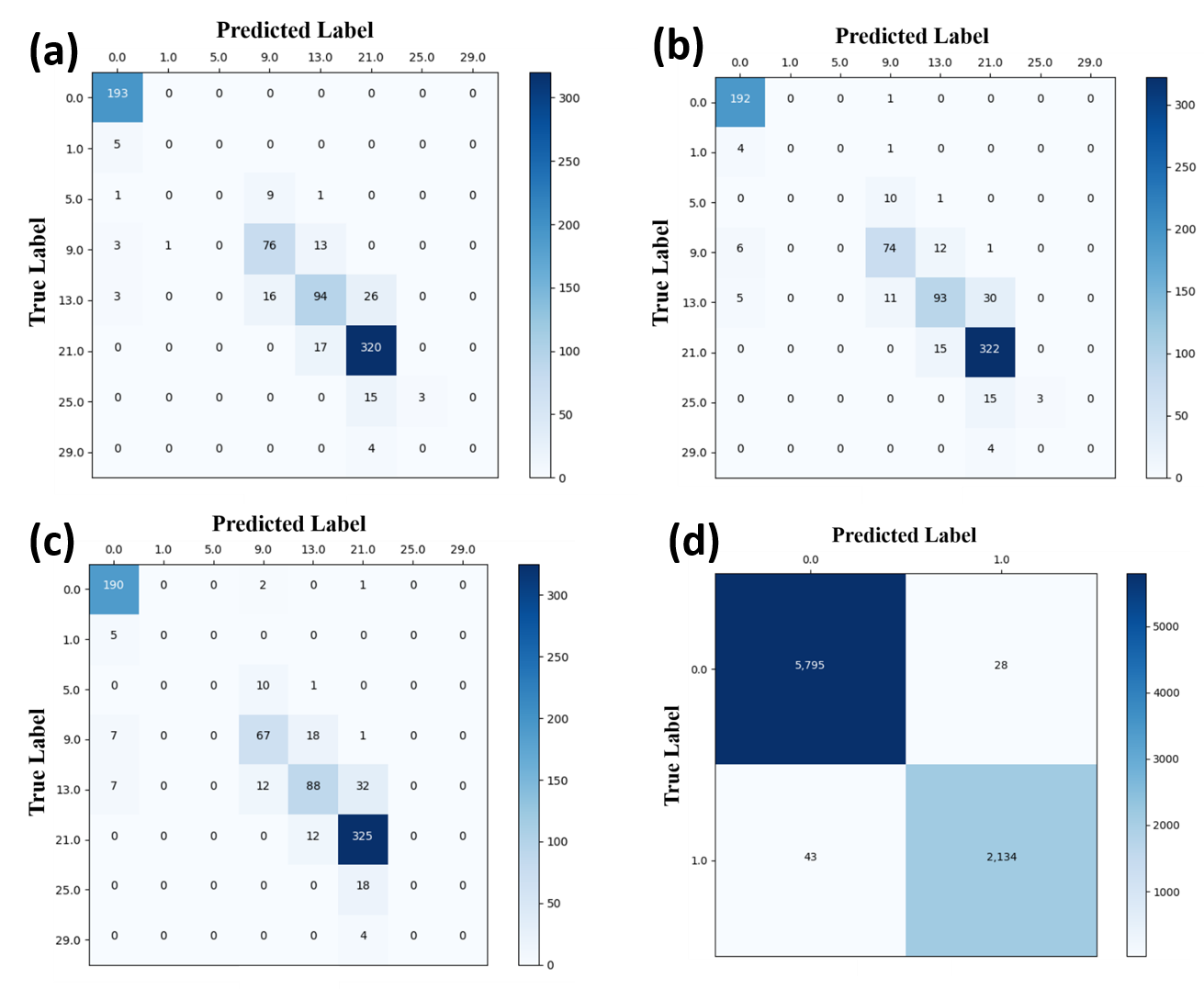}
    \caption{(a) Confusion matrix for classification of intensity of chimera in a 2D lattice network using a trained (a) KNN model (b) SVM model and (c) MLP-NN model. In (a),(b) and (c), numbers on x and y axis correspond to different values of "total number of drifting oscillators" found in a 2D lattice network. (d) Confusion matrix for classification of state of a multilayer network using a trained MLP-NN model. In (d), on the x and y axis, 0 represents a synchronized state and 1 represents a solitary state.}
    \label{confusionMatrix}
\end{figure}

We have seen for a 2D lattice network that MLP-NN is the algorithm best suited to train a machine learning model to predict the exact value of critical delay. Therefore, we use MLP-NN again for generating a prediction model fed on the dataset of the emergent solitary states in multilayer network.
The data is randomly split into training and testing set in the ratio of 4:1. Parameters selected for the neural network are shown in Table~\ref{multilayerMLPParameters}.
We first generated 50 neural network models with  randomly chosen different training sets for each iteration and then the output is averaged for each model to obtain a final value of the critical delay.
The confusion matrix for the MLP-NN model is shown in Fig~\ref{confusionMatrix}d. We also study the sensitivity and the specificity of the MLP-NN model as shown in Table~\ref{multilayerMLPParameters}.
Using the trained multi layer perceptron neural network model, the exact value of the critical delay can be calculated for any pair of interlayer and intra-layer coupling strength (Figs.~\ref{criticalDelay}a,~\ref{criticalDelay}b).

The impact of the threshold for frequency difference to differentiate between solitary state and synchronized state is also studied (see Fig(\ref{criticalDelay}d)). It is observed that if the value of threshold frequency difference is low then changing the threshold value does not have any significant effect on the prediction of the critical delay but as soon as the threshold is changed to a larger value such as a value greater than 0.02 then the machine learning model starts giving wrong predictions.

\begin{table}
\caption{Data structure for Solitary data: There are 8000 rows in total in this table.}
\begin{center}
\scalebox{0.87}{
\begin{tabular}{|c|c|c|c|}
\hline
Interlayer Coupling & Intralayer Coupling & Delay & State \\ \hline
:                   & :                   & :     & :     \\ 
:                   & :                   & :     & :     \\ 
3.60                & 0.86                & 0.50  & 0     \\ 
3.60                & 0.88                & 0.50  & 0     \\ 
3.70                & 0.1                 & 0.50  & 1     \\ 
:                   & :                   & :     & :     \\ 
2.00                & 0.20                & 2.00  & 1     \\ 
2.00                & 0.22                & 2.00  & 0     \\ 
:                   & :                   & :     & :     \\ 
:                   & :                   & :     & :     \\ \hline
\end{tabular}}
\end{center}
\label{solitaryTable}
\end{table}

\begin{table}
\caption{Sensitivity and Specificity of MLP-NN predictive model for multilayer network}
\begin{center}
\scalebox{1}{
\begin{tabular}{|l|l|l|l|}
\hline
\multicolumn{2}{|c|}{Sensitivity} & \multicolumn{2}{c|}{Specificity} \\ \hline
0 (Sync.)               & 1 (Solitary)               & 0 (Sync.)               & 1 (Solitary)              \\ \hline
0.9951          & 0.9802          & 0.9802          & 0.9951         \\ \hline
\end{tabular}
    }
\end{center}
\label{multilayerMLPParameters}
\end{table}

\begin{table}[]
\caption{Parameters for neural network model trained using dataset of multilayer network}
\begin{center}
\scalebox{0.9}{
\begin{tabular}{|l|l|}
\hline
Description                          & Value                  \\ \hline
Total number of layers               & 4                      \\
Total number of hidden layers        & 2                      \\
Number of nodes in each hidden layer & 30                     \\
Number of nodes in the output layer      & 2                      \\
Optimizer                            & Adam                   \\
Learning Rate                        & $10^{-3}$ \\
L2 Penalty                           & $10^{-4}$ \\
Activation                           & ReLU                   \\
Batch Size                           & 200                    \\
Epochs                               & 200                    \\ \hline
\end{tabular}
        }
\end{center}
\label{NNPara2}
\end{table}

\section{Conclusion} 
In this paper, different supervised machine learning algorithms have been employed for the model-free prediction of factors characterizing  the intensity of chimera and solitary states. We demonstrated success of the scheme for two different model systems namely, 2-D lattice and multilayer network. First, chimera states (solitary states) are constructed by instituting delays in the neighboring connections for a selected node (a few isolated interlayer connections) in a 2-D lattice (multiplex network) of Kuramoto oscillators. Next, three machine learning algorithms, K-nearest neighbours, support vector machine and multi-layer perceptron neural network are then put into practice to train the data obtained from the evolution of two network models for the prediction of intensity of rippling chimera states and the value of critical delay to characterize solitary states. It is found that multi-layer perceptron neural network (MLP-NN) classifier makes the most precise prediction in identifying the possible desynchronized oscillators in the rippling chimera states and the value of delay required to tailor the solitary states for a given set of multiplex structural parameters. Furthermore, the trained machine learning model was used to plot the entire phase diagram for the rippling chimera and the solitary state.

To conclude, we demonstrated the success of powerful and model-free machine learning algorithms in tailoring the chimera and the solitary states and anticipate that this investigation would be fruitful in broadening the scope of machine learning techniques in characterizing other dynamical properties and phenomena such as occurrence of explosive synchronization. The study is particularly useful is accessing the systems parameter for experimental setup towards engineering chimeras and Solitary states.

\section*{Acknowledgements}
SJ acknowledges Government of India, CSIR grant 25(0293)/18/EMR-II and DST grant EMR/2016/001921 for financial support. ADK acknowledges CSIR (grant 25(0293)/18/EMR-II), Govt. of India for RA fellowship. SG acknowledges DST, India for INSPIRE (IF150149) fellowship.

\nocite{*}
\bibliography{aipsamp}

\end{document}